\documentclass[10pt,twocolumn,letterpaper]{article}

\usepackage{iccv}              

\definecolor{iccvblue}{rgb}{0.21,0.49,0.74}
\usepackage[pagebackref,breaklinks,colorlinks,allcolors=iccvblue]{hyperref}

\def\paperID{*****} 
\def\confName{ICCV}
\def\confYear{2025}

\title{Towards High-Resolution Alignment and Super-Resolution of Multi-Sensor Satellite Imagery}

\author{Philip Wootaek Shin$^{1,2}$, Vishal Gaur$^3$, Rahul Ramachandran$^2$, Manil Maskey$^2$,\\ Jack Sampson$^1$, Vijaykrishnan Narayanan$^1$, Sujit Roy$^{2,3}$\\
$^1$The Pennsylvania State University \\$^2$NASA Marshall Space Flight Center, Huntsville \\
$^3$University of Alabama in Huntsville 
}
\begin{document}
\maketitle
\begin{abstract}
High-resolution satellite imagery is essential for geospatial analysis, yet differences in spatial resolution across satellite sensors present challenges for data fusion and downstream applications. Super-resolution techniques can help bridge this gap, but existing methods rely on artificially downscaled images rather than real sensor data and are not well suited for heterogeneous satellite sensors with differing spectral, temporal characteristics. In this work, we develop a preliminary framework to align and upscale Harmonized Landsat Sentinel 30m(HLS 30) imagery using Harmonized Landsat Sentinel 10m(HLS10) as a reference from the HLS dataset. Our approach aims to bridge the resolution gap between these sensors and improve the quality of super-resolved Landsat imagery.  Quantitative and qualitative evaluations demonstrate the effectiveness of our method, showing its potential for enhancing satellite-based sensing applications. This study provides insights into the feasibility of heterogeneous satellite image super-resolution and highlights key considerations for future advancements in the field.

\end{abstract}    
\section{Introduction}
\label{sec:intro}

High-resolution remote sensing imagery is essential for various geospatial applications. However, publicly available satellite datasets often provide images at different resolutions and scales, even for the same geographical region. While deep learning-based super-resolution techniques have demonstrated potential in enhancing spatial quality, most existing approaches rely on synthetic low-resolution images generated through conventional downscaling methods such as Lanczos \cite{lanczos} or bicubic interpolation. These models often overlook spatial and temporal variations, including differences in cloud coverage, and assume perfect alignment between high-resolution and low-resolution image pairs, which may not reflect real-world remote sensing conditions. Furthermore, datasets containing heterogeneous satellite images from different sensors that are precisely aligned are limited, with the WorldStrat dataset\cite{WorldStrat} being one of the few available resources.

There has been little research on super-resolving images from distinct satellite sensors with differing native resolutions by leveraging one sensor as a reference to upscale another. In this work, we explore the challenges and opportunities associated with using heterogeneous RGB sensing data for super-resolution.

Contributions we make are three-fold:

\begin{itemize}
\item We develop a preliminary system for aligning and super-resolving satellite images from heterogeneous sensors, specifically using HLS10 as a reference to upscale HLS30. Unlike traditional super-resolution approaches that rely on artificially downscaled images, our method leverages real heterogenous sensor data with distinct resolutions and characteristics.
\item We incorporate Histogram Matching (HM)\cite{HM} and Feature Distribution Matching (FDM)\cite{abramov2020keep} as preprocessing steps to mitigate spectral discrepancies between heterogeneous sensors before applying super-resolution refinement. This improves the performance of diffusion-based refinement models, such as SR3, which otherwise struggle with cross-sensor variations.

\item We curate a custom dataset of co-registered multi-spectral imagery from HLS 30 and HLS 10 across 18 diverse terrain regions with minimal cloud coverage. This dataset facilitates multi-channel super-resolution beyond RGB, incorporating SWIR1, SWIR2, and NIR channels, and enables exploration of training on surface reflectance values which are calculated from the raw digital numbers captured by the satellite sensors, for advancing multi-spectral super-resolution techniques.
\end{itemize}
\section{Related Work}
\label{sec:RelatedWork}
We provide a brief overview of three pertinent research areas. First, we discuss satellite image upscaling, highlighting the approaches that inspired our methodology. Next, we introduce satellite datasets, with a focus on the datasets used for testing in our study. Finally, we describe our image alignment strategy, which addresses the challenges of processing heterogeneous sensor data from different satellite sources.

\subsection{Remote Sensing Upscaling}
\label{ssec:remotesensing}

Remote Sensing Super-Resolution (RSSR) focuses on high-resolution (HR) reconstruction from one or more low-resolution (LR) images, facilitating critical tasks such as object detection and semantic segmentation in satellite imagery\cite{ediffsr, liu2022diffusion, xu2023dual, ali2023tesr}. Additionally, pan-sharpening techniques, widely used in remote sensing, aim to fuse panchromatic images (high spatial resolution but low spectral resolution) with multispectral images (lower spatial resolution but higher spectral resolution) to produce HR multispectral images, thus enhancing spatial and spectral details\cite{yang2023panflownet, rui2024unsupervised,li2023hyperspectral}. After empirical trials, we selected the TESR\cite{ali2023tesr} as our backbone architecture. TESR’s combination of Vision Transformer-based and diffusion-based refinement stages makes it particularly well-suited for addressing the challenges posed by heterogeneous data.



\subsection{Dataset}
\label{ssec:dataset}
Remote sensing datasets play a crucial role in advancing research and applications in satellite imagery analysis. Among the most widely used datasets are Landsat and Sentinel, which provide multispectral imagery for a variety of Earth observation tasks\cite{hls_dataset}. The WorldStrat\cite{WorldStrat} dataset represents a significant step forward in providing aligned imagery from different sensors. Unlike Landsat and Sentinel, which focus on continuous data acquisition from a single sensor type, WorldStrat includes images from heterogeneous sensors (e.g., SPOT-6/7 and Sentinel-2) covering the same geographical regions. 

Harmonized Landsat‑8 and Sentinel‑2 (HLS)\cite{hls_dataset} surface reflectance dataset is a NASA initiative to create a virtual constellation combining Landsat‑8 and Sentinel‑2 products, harmonized to surface reflectance with consistent atmospheric correction, BRDF normalization, spatial co‑registration, and band‑matching\cite{Nguyen2020HarmonizationCropMonitoring},\cite{Claverie2018HLS} 
\subsection{Image alignment}
\label{ssec:imagealign}

Conventional image alignment methods, such as Histogram Matching (HM)\cite{HM}, are widely used in image processing to establish a monotonic mapping between the histograms of two images, ensuring their intensity distributions are aligned. Additionally, deep learning techniques\cite{li2024color,yu2018deepexposure} have been developed to address more complex alignment tasks, such as correcting overexposed or underexposed images by learning non-linear transformations. Another prominent approach is statistical matching\cite{abramov2020keep}, which uses statistical metrics to align images by matching their distributions or features, making it particularly effective for scenarios involving complex variations between datasets.

\subsection{Surface reflectance}
\label{ssec:surfacereflectance}

To convert Digital Number (DN) to surface reflectance, we first compute the spectral radiance:

\begin{equation}
L_\lambda = G \cdot \text{DN} + B
\end{equation}

where $L_\lambda$ is the spectral radiance (in W\,m$^{-2}$\,sr$^{-1}$\,$\mu$m$^{-1}$), $G$ is the radiometric gain, $B$ is the bias (offset), and DN is the digital number from the sensor.

Top-of-atmosphere (TOA) reflectance is then computed as:

\begin{equation}
\rho_{\text{TOA}} = \frac{\pi \cdot L_\lambda \cdot d^2}{E_{\text{sun},\lambda} \cdot \cos(\theta_s)}
\end{equation}

where $\rho_{\text{TOA}}$ is the TOA reflectance (unitless), $d$ is the Earth–Sun distance in astronomical units, $E_{\text{sun},\lambda}$ is the mean solar exoatmospheric irradiance, and $\theta_s$ is the solar zenith angle.

An optional atmospheric correction yields surface reflectance:

\begin{equation}
\rho_{\text{surf}} = \frac{\rho_{\text{TOA}} - \rho_{\text{path}}}{T_s \cdot T_v}
\end{equation}

where $\rho_{\text{surf}}$ is the surface reflectance, $\rho_{\text{path}}$ is the atmospheric path reflectance, and $T_s$ and $T_v$ are the atmospheric transmittance from the sun to the surface and from the surface to the sensor, respectively.

 Beyond multi-sensor fusion, surface reflectance values are foundational in numerous geoscience applications. For example, they are widely used in vegetation monitoring through indices such as Normalized Difference Vegetation Index(NDVI) and Enhanced Vegetation Index(EVI) \cite{Didan2015MOD13}, in land cover and land use classification \cite{Zhang2021LandCover}, and in surface water and flood mapping using indices like Normalized Difference Water Index(NDWI) and Modified Normalized Difference Water Index(MNDWI) \cite{Xu2006NDWI}. Additionally, surface reflectance is vital in soil and mineral mapping, where the spectral reflectance curves help distinguish between different surface materials \cite{BenDor2009HyperspectralSoil}. In climate studies, accurate surface reflectance is used to estimate surface albedo, which influences energy balance models and carbon flux estimations \cite{Schaaf2002MODISAlbedo}. 

Because surface reflectance captures intrinsic surface properties independent of atmospheric and illumination conditions, it serves as a robust and physically meaningful supervisory signal for training deep learning models. Using surface reflectance in the loss function ensures that the network optimizes for values directly relevant to geophysical processes, enabling better generalization across time, geography, and sensors.
\section{Method}

\label{sec:Method}

\begin{figure}[t]
    \begin{center}
        \includegraphics[width=0.8\linewidth]{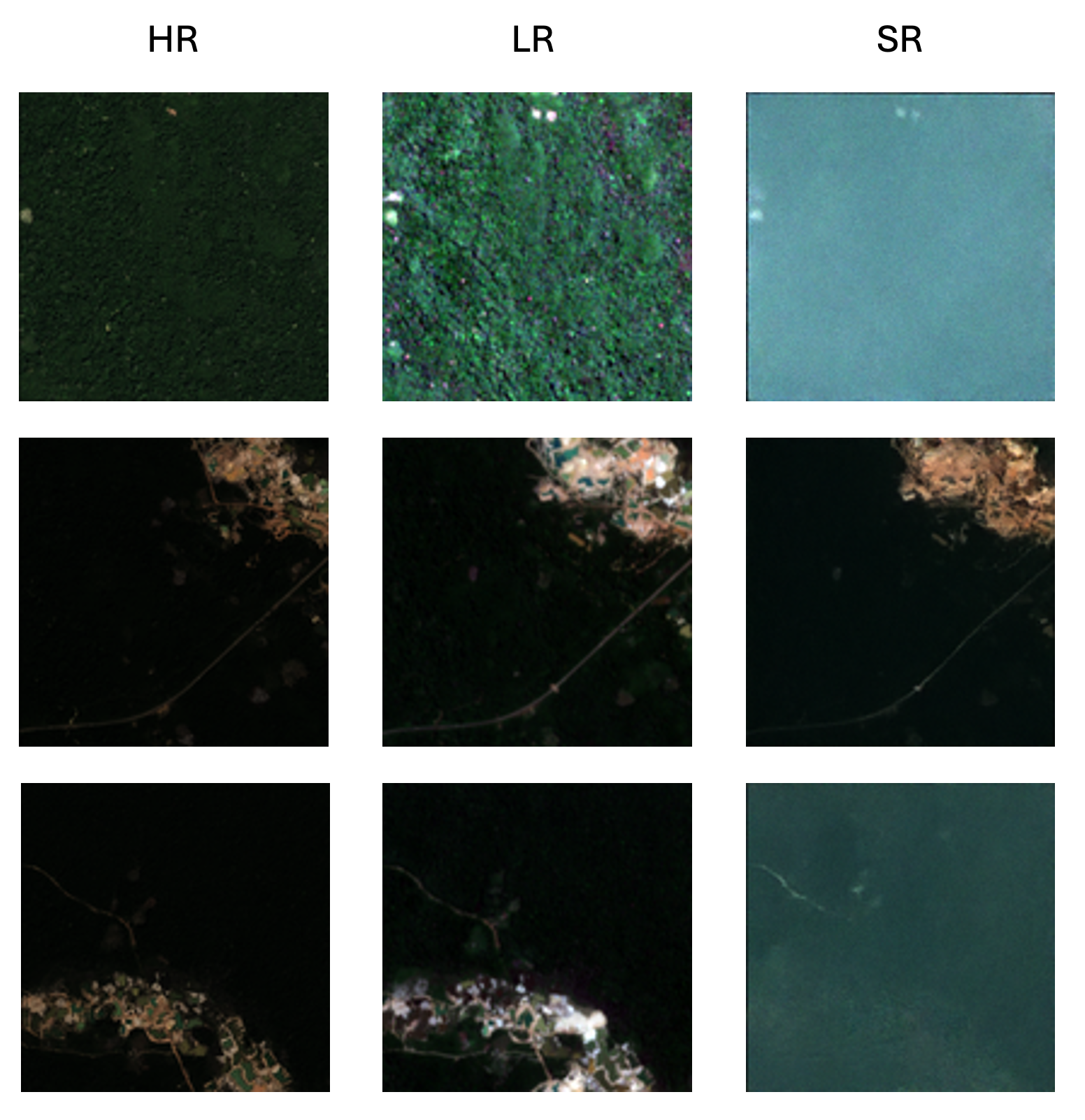}
    \end{center}
\caption{Example of a high-resolution (HR) image, a low-resolution (LR) image from the WorldStrat dataset, and the super-resolved (SR) image generated using the two-stage TESR architecture. The refinement step in the second stage did not perform effectively, leading to suboptimal super-resolution results, as evident in the visual comparison.}
\label{fig:fig1}
\end{figure}


For this study, we utilized the WorldStrat dataset \cite{WorldStrat} along with a custom dataset. WorldStrat provides high-resolution RGB imagery from SPOT-6/7 (1.5m per pixel) and low-resolution multispectral imagery from Sentinel-2 (10m per pixel). However, discrepancies in image size ratios were observed, with high-resolution images (1054 × 1054 pixels) and low-resolution images (140–150 pixels per side) not aligning perfectly in pixel-per-meter scaling. Despite covering the same geographic regions, these inconsistencies posed challenges for direct comparisons and processing.


For consistency, we downscaled high-resolution images to 384 × 384 pixels and low-resolution RGB images to 128 × 128 pixels using the Lanczos method \cite{lanczos}. This preprocessing aligns with our goal of applying 3× magnification to enhance Harmonized Landsat Sentinel per pixel resolution.

In dataset, the high-resolution (SPOT-6/7) and low-resolution (Sentinel-10) images originate from heterogeneous sensors and satellites, leading to inherent differences in their characteristics


\subsection{Baseline}
\label{ssec:Baseline}

Since the high-resolution images of Worldstrat dataset were in PNG format with an RGB range of 0–255, we converted the low-resolution Sentinel-10 images into the same PNG format and RGB range for consistency in preliminary experiments. After empirical experiments, we selected the TESR architecture \cite{ali2023tesr} for our approach. TESR leverages the combined strengths of Vision Transformers and Diffusion Models to artificially enhance the resolution of remote sensing (RS) images.

The TESR architecture operates in two stages. The first stage utilizes a Vision Transformer-based model, SwinIR \cite{Swinir}, to increase image resolution. The second stage employs an iterative Diffusion Model, SR3 \cite{SR3}, pre-trained on a larger dataset, to enhance the quality of the generated high-resolution images. This two-stage process is designed to achieve both resolution enhancement and visual refinement.


\begin{figure}[!t]
    \begin{center}
        \includegraphics[width=0.9\linewidth]{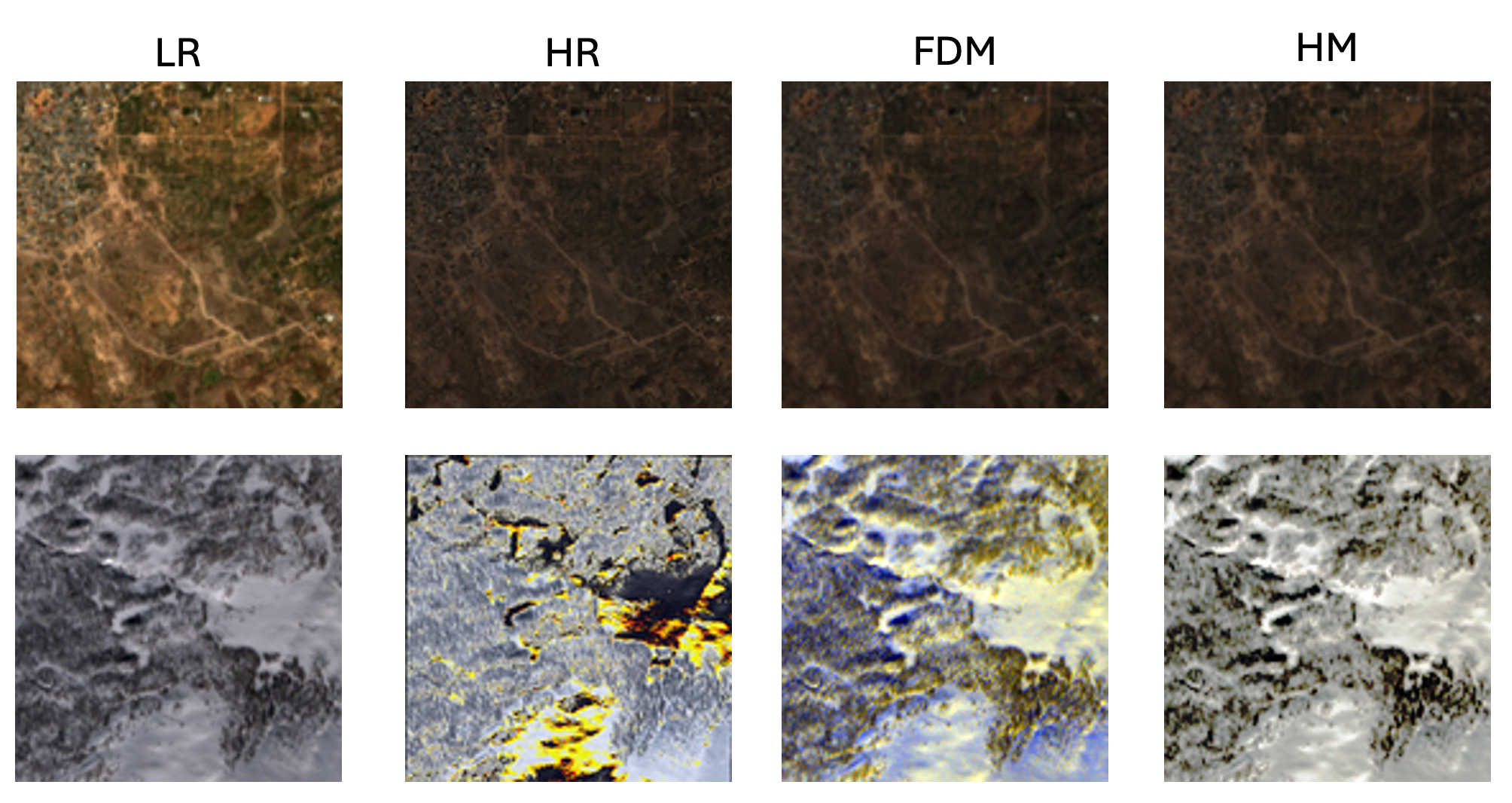}
    \end{center}
\caption{Illustration of low-resolution (LR) and high-resolution (HR) images, alongside results after applying Histogram Matching (HM) and Feature Distribution Matching (FDM). In the top example, there is minimal visual difference between the HM and FDM outputs. In contrast, the bottom example shows that while both HM and FDM outputs are somewhat closer to the reference HR image, neither aligns as well as observed in the top example}
\label{fig:fig2}
\end{figure}

\begin{figure*}[t]
\begin{center}
\includegraphics[width=0.85\linewidth]{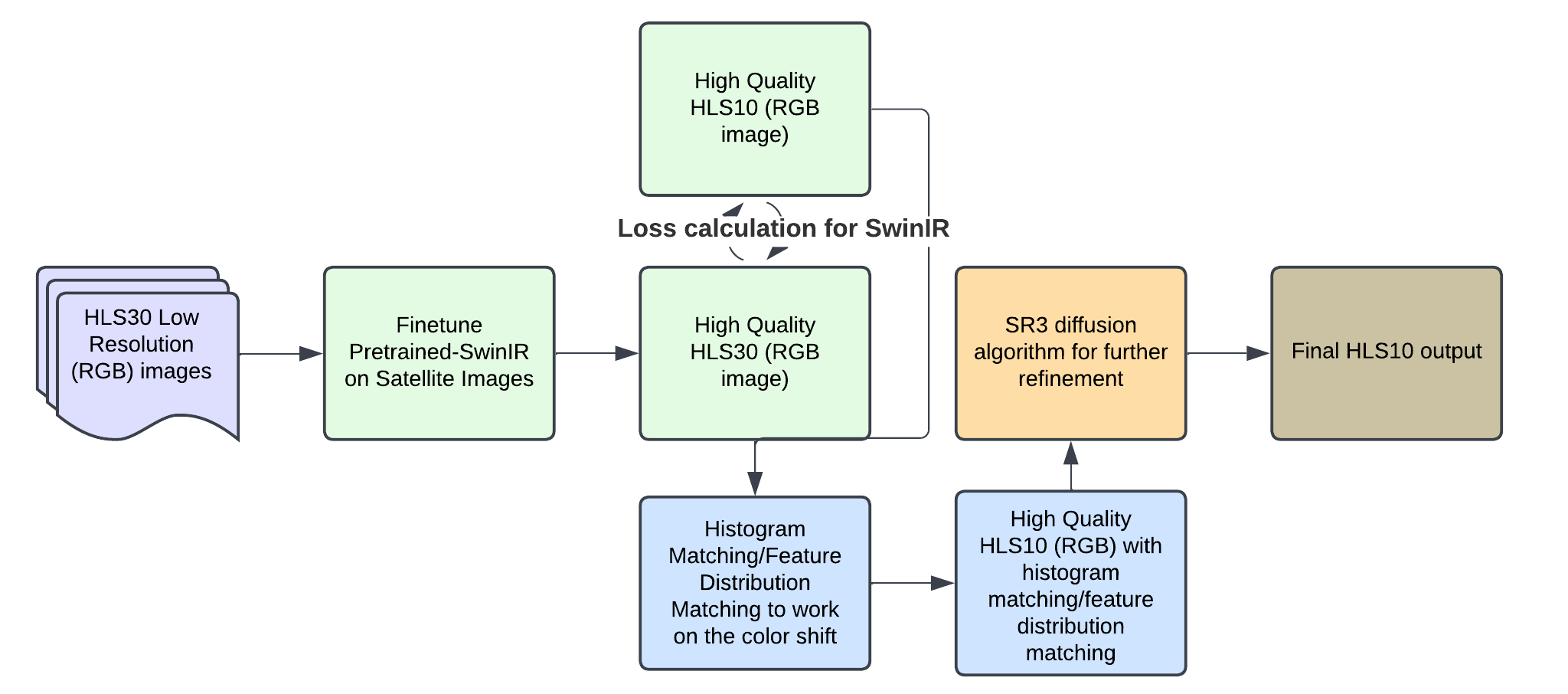}
\end{center}
\caption{The full pipeline for 3 channel RGB remote sensing data}
\label{fig:fullpipeline}
\end{figure*}

When applying the two-stage TESR pipeline, the first stage using SwinIR successfully upscales the low-resolution image to match the size of the high-resolution image. However, it does not directly reproduce the high-resolution image due to differences in histogram distributions caused by the distinct sensors from which the images originate. Additionally, issues arise in the second stage, where the refinement process struggles to perform effectively, further exacerbated by the sensor discrepancies. As shown in Fig.~\ref{fig:fig1}, the refinement stage produces suboptimal results, characterized by visible artifacts and poor training outcomes. 

These limitations hinder the ability of the pipeline to accurately refine and enhance the details in the upscaled images. While the images cover the same geographic regions and are structurally similar, discrepancies in color and visual features are apparent due to the distinct properties of the sensors. This heterogeneity might significantly impact network training and necessitate a specialized approach for super-resolution.

\subsection{Method Proposed}
\label{ssec:MethodProposed}

To address the issue of color shift, we explored two methods: Histogram Matching(HM)\cite{HM} and Feature Distribution Matching(FDM)\cite{abramov2020keep}. Feature Distribution Matching(FDM) is built on the color transfer method proposed by Xiao and Ma \cite{10.1145/1128923.1128974}. This approach aims to transform a source image to acquire the color mean and covariance of a target image while preserving the content of the source image. Unlike traditional transformations in homogeneous coordinates, FDM generalizes the transformation to a c-dimensional Euclidean space, providing a more flexible and robust solution for aligning feature distributions between heterogeneous datasets.

The visual and quantitative results of applying Feature Distribution Matching (FDM) and Histogram Matching (HM) are presented in Fig.~\ref{fig:fig2}. The figure demonstrates that both FDM and HM show potential as solutions to improve the TESR pipeline's second stage (SR3-based refinement).

To evaluate the qualitative performance of HM and FDM, we conducted an analysis on a subset of the WorldStrat dataset. Out of a total of 3927 images, 1\% of the data (39 randomly selected images) was used for testing. Initially, SwinIR was applied to upscale the low-resolution images to match the size of the high-resolution images. Subsequently, FDM and HM were applied to these upscaled images.

For quantitative analysis, we measured PSNR, LPIPS\cite{lpips}, and SSIM\cite{ssim} metrics for the 39 selected images. The results, summarized in Tab.~\ref{tab:QuantHMFDM}, indicate that both FDM and HM outperform the baseline of simply upscaling low-resolution images without applying FDM or HM. Moreover, while FDM excelled in certain metrics, HM performed better in others, suggesting that the choice between FDM and HM may depend on the specific metric or application context.

\begin{table}[t]
    \centering
    \footnotesize
    \begin{tabular}{lccc}
        \hline
        \textbf{Method} & \textbf{SSIM($\uparrow$)} &  \textbf{LPIPS($\downarrow$)} &\textbf{PSNR($\uparrow$)} \\
        \hline
        HR LR(SR) & 0.5277 & 0.2868 & 19.1868 \\
        HR FDM & 0.6330  & \textbf{0.2020} & \textbf{28.9380}\\
        HR HM & \textbf{0.6639}  & 0.2152 & 28.8626\\
        \hline
    \end{tabular}
  
    \caption{Comparison of three methods with three metrics: (1) HR image with LR upscaled using SwinIR, (2) HR image with LR upscaled using SwinIR followed by Feature Distribution Matching (FDM), and (3) HR image with LR upscaled using SwinIR followed by Histogram Matching (HM).}
    \label{tab:QuantHMFDM}
\end{table}

The complete pipeline is illustrated in Fig.~\ref{fig:fullpipeline}. The process begins with inputting the low-resolution (LR) image, which is upscaled using SwinIR to match the size of the high-resolution (HR) image, resulting in a super-resolved image (SR). Next, FDM and HM are applied to address color shifts and align the pixel distribution of the SR image with that of the HR image. Finally, the refined SR image undergoes further enhancement using the SR3 refinement stage.

To train the SR3 network for refinement, we utilized 99\% of the WorldStrat dataset, comprising 3888 images, with the remaining 1\% (39 images) reserved for validation to evaluate the metrics. The training process was conducted over 300,000 iterations (610 epochs) with a batch size of 8. For inference on the validation set, we used 2000 iterations to generate the output images. The qualitative analysis is detailed in Tab.~\ref{tab:fullpipe}.

\begin{table}[t!]
    \centering
    \footnotesize
    \begin{tabular}{lccc}
        \hline
        \textbf{Method} & \textbf{SSIM($\uparrow$)} &  \textbf{LPIPS($\downarrow$)} &\textbf{PSNR($\uparrow$)} \\
        \hline
        HR XHMSR & 0.1276 & 0.5218 & 10.8132 \\
        HR FDMSR & \textbf{0.5268}  & 0.2359 & \textbf{27.6825}\\
        HR HMSR & \textbf{0.6181}  & \textbf{0.2232} & 27.4161\\
        \hline
    \end{tabular}
  
    \caption{Comparison of three methods evaluated using three metrics: (1) HR image with LR upscaled using SwinIR, followed by SR3 refinement without applying Histogram Matching (HM) or Feature Distribution Matching (FDM); (2) HR image with LR upscaled using SwinIR, followed by FDM and SR3 refinement; and (3) HR image with LR upscaled using SwinIR, followed by HM and SR3 refinement.}
    \label{tab:fullpipe}
\end{table}

\subsection{Custom dataset}
\label{ssec:CustomDataset}

For our custom dataset, we collected 30-meter resolution HLS tiles from the CMR EarthData search spanning 2018 to 2022. These tiles were processed through the Harmonized Landsat and Sentinel-2 (HLS) pipeline to ensure consistency in atmospheric correction, BRDF normalization, spatial co-registration, and band alignment. We refer to this Landsat-derived 30-meter HLS product as HLS30/L30 throughout the paper. In contrast, the 10-meter resolution tiles were sourced from a private NASA GSFC repository. Throughout this study, we refer to the 10-meter resolution tiles as HLS10 or S10, which specifically denote Sentinel-2 imagery that has undergone the full HLS processing pipeline, including atmospheric correction, BRDF normalization, spatial co-registration, and band-matching. As Landsat satellites do not provide native 10-meter resolution data, all references to HLS10/S10 pertain exclusively to Sentinel-2.

The images fetched had a 100\% spatial coverage for 10m resolution and 30m resolution tiles, ensuring less than 5\% cloud coverage. The HLS30 tiles had dimensions of 3660 $\times$ 3660 pixels, while the HLS10 tiles measured 10980 $\times$ 10980 pixels. Both datasets utilized six spectral bands (SWIR1, SWIR2, NIR, Red, Green, and Blue), stored as single TIFF files.



For consistency, HLS30 images were cropped into 128 × 128 pixel patches and HLS10 images into 384 × 384 pixel patches from the upper-left corner, preserving metadata. Any excess pixels were discarded, resulting in 28 × 28 patches per terrain image, yielding 784 patches per dataset image.

Statistical analysis was performed on the patches of both both resolution tiles/images to calculate the distribution, minimum, maximum, mean, and standard deviation of the dataset. These values were calculated for both, the overall dataset and each individual band. The minimum and maximum values were used for min-max normalization as a pre-processing step within the data loader before passing the data into the model.

\section{Experiements}
\label{sec:Experiement}

\begin{figure}[t]
    \begin{center}
        \includegraphics[width=0.6\linewidth]{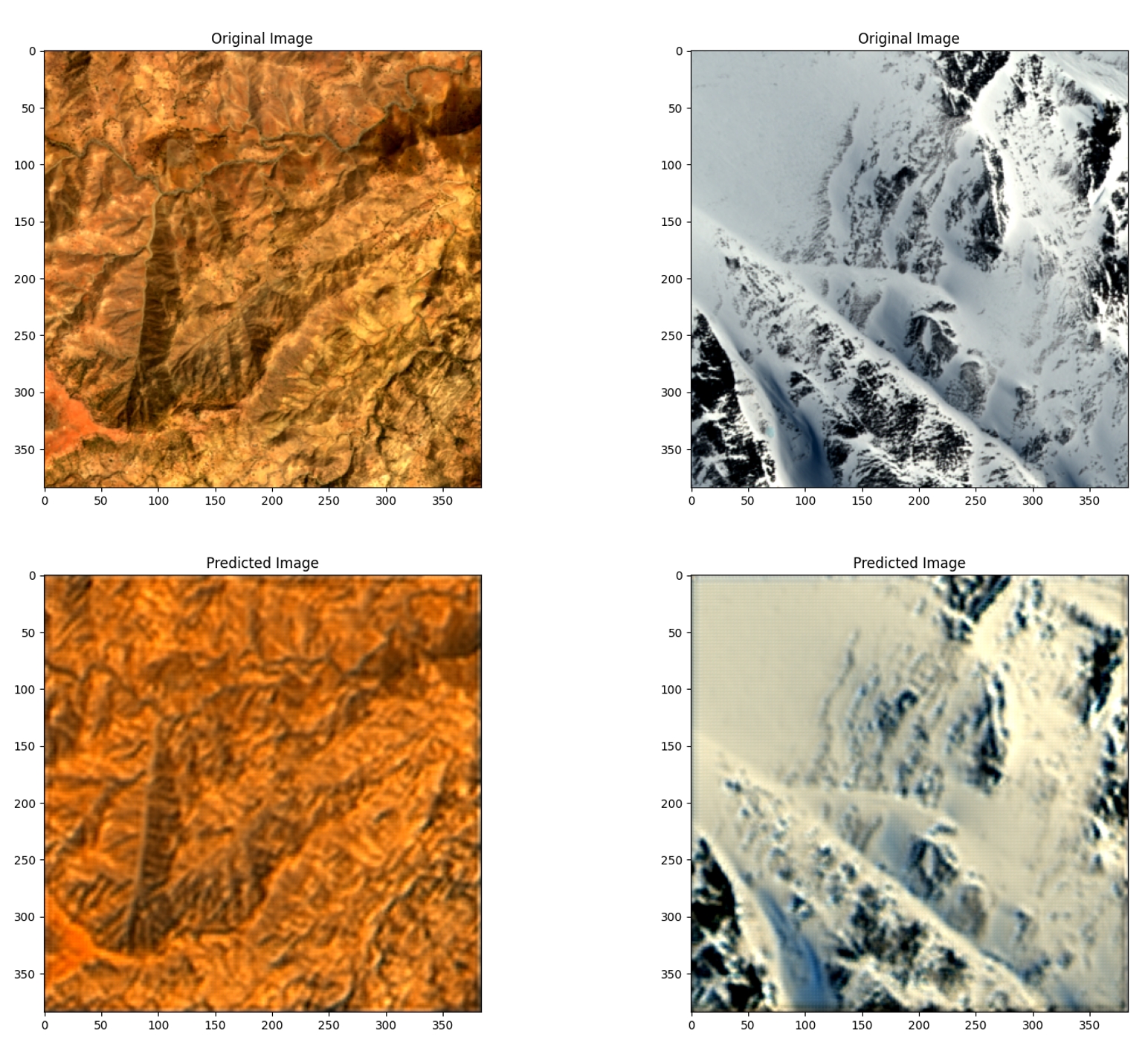}
    \end{center}
\caption{Illustration of the original image and the predicted image, trained using transfer learning with SwinIR on surface reflectance value and clamped to PNG format for visualization. The results indicate that while transfer learning with SwinIR facilitated some level of feature adaptation, finer details were not captured with sufficient precision, and noticeable color shifts occurred in the predicted image.}
\label{fig:fig4}
\end{figure}

In the context of the WorldStrat dataset, our experiments were limited to RGB channels, with pixel values restricted to the range of 0–255. This is not representative of typical remote sensing data, where pixel values often range from -3000 to 14000. 


To conduct experiments with geoscientific relevance, we converted the sensor measurements i.e., digital numbers into surface reflectance values. To adapt the SwinIR architecture for our dataset, we utilized its six Residual Swin Transformer Blocks (RSTBs), freezing the first three blocks and fine-tuning the last three using three-channel RGB surface reflectance data. Additionally, we modified the network architectures of both SR3 and SwinIR to support inputs with more than three channels and surface reflectance values. However, for the scope of this study, we focus exclusively on the RGB channels.

\begin{figure*}[t]
    \begin{center}
        \includegraphics[width=0.87\linewidth]{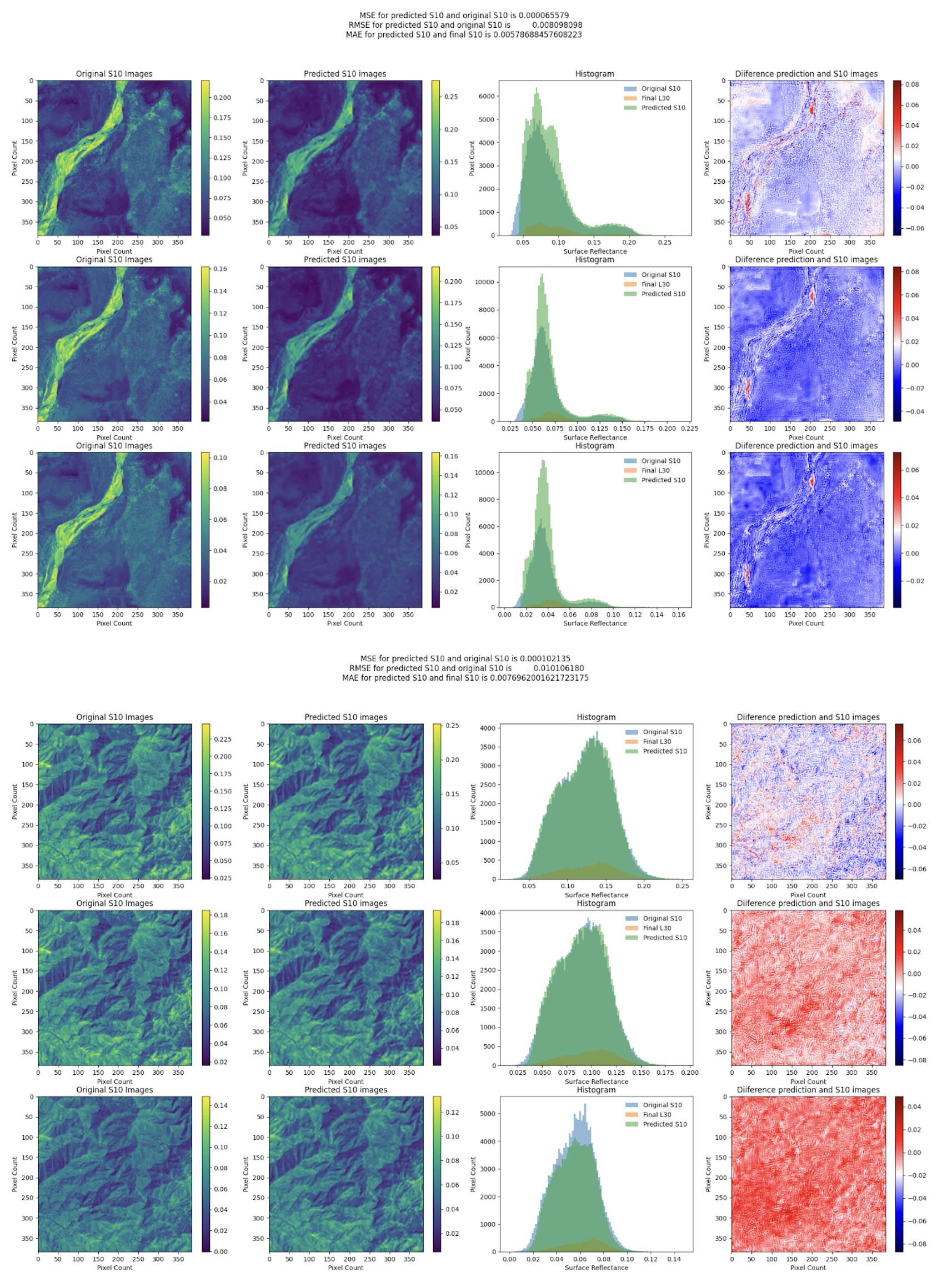}
    \end{center}
\caption{Illustration of qualitative and quantitative comparisons of surface reflectance values across red, green, and blue bands between original HLS10(S10), predicted HLS10(S10), and HLS30(L30) inputs.  }
\label{fig:fig5}
\end{figure*}

\subsection{Finetuning SwinIR with Remote Sensing Data}
\label{ssec:OurPipeline}
The rationale behind this approach is rooted in the hierarchical learning paradigm typical of neural networks. The earlier layers are generally responsible for capturing high-level structural patterns and relationships in the input data, while the later layers focus on learning finer details and task-specific features. By freezing the initial layers, we preserved the model's ability to capture general structural features, while fine-tuning the later layers allowed the network to adapt to the unique characteristics of the custom dataset.

Although the pretrained weights of SwinIR are derived from training on natural images, they provide a significant advantage when compared to training from scratch on a restricted hyperspectral dataset. Natural image datasets are vast and diverse, enabling the pretrained weights to capture robust, generalized features such as edges, textures, and structural patterns that are often present across different types of imagery. While hyperspectral images have unique characteristics, particularly in their spectral dimensions, the spatial features learned from natural images can still serve as a strong foundation. Transfer learning allows the network to leverage this prior knowledge, reducing the training time and computational cost while achieving better performance on the target task. Additionally, given that hyperspectral datasets are often limited in size and variety, using pretrained weights helps mitigate overfitting and improves the network's ability to generalize, making it an efficient and effective choice for adapting to hyperspectral data.

Our result for transfer learning SwinIR is presented in Fig.~\ref{fig:fig4}. As shown in figure the original image is the HR image from S10 and the predicted image is the inferenced image that were trained on surface reflectance value for SwinIR. We can still observe the color shift and the finer details are not preserved well with the SwinIR like we expected so we applied the following Histogram Matching or Feature Distribution matching and SR3 training for finer details


\subsection{Histogram Matching/Feature Distribution Matching with SR3}
\label{ssec:SR3}

Instead of employing deep learning methods for image alignment, we opted for Histogram Matching (HM) and Feature Distribution Matching (FDM) due to their efficiency and seamless integration with the SR3 and SwinIR pipelines. By incorporating HM/FDM as a single loss function, we avoided the complexity of using multiple models for the task. Additionally, these methods delivered satisfactory performance, eliminating the need for more computationally intensive deep learning approaches.

For training the SR3, channel-wise normalization was employed instead of global normalization. During experimentation, global normalization across all channels led to suboptimal results for the inferred/generated super-resolved (SR) images, likely due to the varying characteristics of each channel in the sensor data.




\section{Results}
\label{sec:Results}


Fig.\ref{fig:fig5} presents a comparative visualization comprising six rows and four columns. The rows are organized by spectral channels, where the first and fourth rows correspond to the red channel, the second and fifth rows to the green channel, and the third and sixth rows to the blue channel of the sensor. The first column displays the reference surface reflectance values from the original HLS10 sensor. The second column shows the predicted HLS10 reflectance values obtained through upscaling from the HLS10 sensor input. The third column provides histograms comparing the pixel distributions of the original HLS10, the predicted HLS10, and the final L30 reflectance values. The fourth column illustrates the difference maps between the original HLS10 and the predicted HLS10 images, highlighting spatial discrepancies in surface reflectance.

We included bandwise image comparisons to facilitate more targeted analysis by geoscientists, enabling a detailed interpretation of how individual spectral bands contribute to the reconstructed outputs. While RGB composite images provide an intuitive overview, a deeper dive into individual bands of the S10 product reveals how each channel is affected differently by the upscaling process. This analysis also highlights the model’s band-specific performance, shedding light on potential spectral inconsistencies that may be overlooked in aggregate metrics.  This framework can be extended to identify and mitigate inefficiencies specific to individual bands, guiding more robust and interpretable model development for geoscientific tasks.

In addition to the visual comparison, we report quantitative evaluation using Mean Squared Error (MSE), Root Mean Squared Error (RMSE), and Mean Absolute Error (MAE) to assess the fidelity of the predicted HLS10 images. These pixel-wise metrics were selected because conventional perceptual quality metrics such as LPIPS, SSIM, and PSNR are often less effective in capturing subtle yet critical differences in surface reflectance, especially in scientific imaging tasks. Our approach prioritizes numerical accuracy over perceptual similarity, making these error metrics more appropriate for evaluating the physical consistency of the reconstructed images.

As illustrated in the histograms presented in the third column, the distribution of surface reflectance values for the final HLS30 and original HLS10 images differs, despite both capturing the same terrain. This discrepancy arises due to inherent differences between the HLS30 and HLS10 sensors. Additionally, because the HLS30 sensor has a coarser spatial resolution (30 meters) compared to the S10 sensor (10 meters), the number of pixels in the HLS30(L10) data is approximately one-third that of the S10, resulting in a lower pixel count.

Across all red, green, and blue channels for the two example regions, the predicted HLS10 distributions closely align with those of the original S10 images, whereas the HLS30 distributions deviate more noticeably. This alignment supports the effectiveness of our model architecture in accurately upscaling HLS30 sensor data to produce high-resolution HS10-level predictions. To further assess the prediction accuracy, the fourth column visualizes the pixel-wise difference between the predicted and original HLS10 images. These difference maps indicate that the discrepancies in surface reflectance values are minimal, demonstrating that our model preserves spatial and spectral fidelity during the upscaling process. 


\section{Conclusion}
\label{sec:Conclusion}

This paper makes three key contributions to the field of remote sensing super-resolution. First, we present a preliminary system for aligning and super-resolving satellite images from heterogeneous sensors, HLS 30 HLS 10. Second, we incorporate Histogram Matching (HM) and Feature Distribution Matching (FDM) as preprocessing steps to address spectral discrepancies between heterogeneous sensors, improving the effectiveness of super-resolution refinement. Third, we collect and curate a custom dataset consisting of co-registered multi-spectral imagery from HLS30 and HLS10, spanning 18 terrain regions with minimal cloud coverage, to facilitate research in this domain. Looking ahead, our goal is to expand this framework to align and upscale heterogeneous sensor data using channels beyond RGB, further enhancing the applicability of super-resolution techniques in multi-spectral remote sensing.

\begin{center}
\section*{Acknowledgements}
\end{center}

This work was supported by NASA Grant 80MSFC22M004, in part by Semiconductor Research Corporation JUMP 2.0 PRISM Center and NSF Award 2243979.

{
    \small
    \bibliographystyle{ieeenat_fullname}
    \bibliography{main}
}

\end{document}